\pdfoutput=1
\PassOptionsToPackage{x11names}{xcolor}
\documentclass[11pt]{article}

\usepackage[T1]{fontenc}
\usepackage[sc]{mathpazo}
\usepackage{fullpage}

\usepackage{color}
\usepackage{amsmath,amssymb,amsfonts,amsthm}
\usepackage{comment}
\usepackage{xspace}
\usepackage{authblk}

\usepackage[colorlinks=true,citecolor=Blue4,linkcolor=Firebrick3]{hyperref}

\usepackage[nameinlink]{cleveref}

\usepackage{pgf,tikz}
\usetikzlibrary{arrows,automata,decorations.pathmorphing,fit,positioning,arrows.meta,calc}

\tikzstyle{mynode}=[draw,circle,minimum size=1cm,inner sep=0pt,thick,fill=white,font=\Large]
\tikzstyle{myedge}=[-{Latex[length=2mm, width=2mm]}, thick]
\tikzstyle{mysnake}=[decorate,decoration={snake,amplitude=.4mm,segment length=2mm,post length=1mm}]

\newcommand{\convexpath}[2]{
[   
    create hullnodes/.code={
        \global\edef\namelist{#1}
        \foreach [count=\counter] \nodename in \namelist {
            \global\edef\numberofnodes{\counter}
            \node at (\nodename) [draw=none,name=hullnode\counter] {};
        }
        \node at (hullnode\numberofnodes) [name=hullnode0,draw=none] {};
        \pgfmathtruncatemacro\lastnumber{\numberofnodes+1}
        \node at (hullnode1) [name=hullnode\lastnumber,draw=none] {};
    },
    create hullnodes
]
($(hullnode1)!#2!-90:(hullnode0)$)
\foreach [
    evaluate=\currentnode as \previousnode using \currentnode-1,
    evaluate=\currentnode as \nextnode using \currentnode+1
    ] \currentnode in {1,...,\numberofnodes} {
-- ($(hullnode\currentnode)!#2!-90:(hullnode\previousnode)$)
  let \p1 = ($(hullnode\currentnode)!#2!-90:(hullnode\previousnode) - (hullnode\currentnode)$),
    \n1 = {atan2(\y1,\x1)},
    \p2 = ($(hullnode\currentnode)!#2!90:(hullnode\nextnode) - (hullnode\currentnode)$),
    \n2 = {atan2(\y2,\x2)},
    \n{delta} = {-Mod(\n1-\n2,360)}
  in 
    {arc [start angle=\n1, delta angle=\n{delta}, radius=#2]}
}
-- cycle
}

\newtheorem{theorem}{Theorem}[section]
\newtheorem{lemma}[theorem]{Lemma}
\newtheorem{problem}[theorem]{Problem}
\newtheorem{observation}[theorem]{Observation}
\newtheorem{remark}[theorem]{Remark}
\newtheorem{corollary}[theorem]{Corollary}

\crefname{theorem}{Theorem}{Theorems}
\crefname{lemma}{Lemma}{Lemmas}
\crefname{claim}{Claim}{Claims}
\crefname{remark}{Remark}{Remarks}
\crefname{observation}{Observation}{Observations}
\crefname{corollary}{Corollary}{Corollaries}
\crefname{appendix}{Appendix}{Appendices}
\crefname{section}{Section}{Sections}
\crefname{equation}{Eq.}{Eqs.}
\crefname{algorithm}{Algorithm}{Algorithms}
\crefname{figure}{Figure}{Figures}
\crefname{table}{Table}{Tables}

\newcommand{\bigO}{\mathcal{O}}
\newcommand{\RU}{R^\bigtriangledown}
\newcommand{\RD}{R_\bigtriangleup}
\newcommand{\SCC}{\mathrm{SCC}}

\usepackage{algorithmicx}
\usepackage[ruled,section]{algorithm}
\usepackage[noend]{algpseudocode}
\algnewcommand\And{\; \textbf{and} \;}
\algnewcommand\Or{\; \textbf{or} \;}
\algnewcommand\To{\; \textbf{to} \;}
\algnewcommand\Continue{\textbf{continue}}
\algnewcommand\Not{\textbf{not}}

\algrenewcommand\textproc{\textsl}

\title{A Fully Polynomial Parameterized Algorithm for Counting the Number of Reachable Vertices in a Digraph}
\author{Naoto Ohsaka\thanks{\href{mailto:naoto.ohsaka@gmail.com}{\texttt{naoto.ohsaka@gmail.com}}}}
\affil{NEC Corporation}
\date{\vspace{-5ex}}

\begin{document}

\maketitle

\begin{abstract}

We consider the problem of counting the number of vertices reachable from each vertex in a digraph $G$,
which is equal to computing all the out-degrees of the transitive closure of $G$.
The current (theoretically) fastest algorithms run in quadratic time; however,
Borassi has shown
that this problem is not solvable in truly subquadratic time
unless the Strong Exponential Time Hypothesis fails
[Inf.~Process.~Lett., 116(10):628--630, 2016].
In this paper, we present an $\bigO(f^3n)$-time exact algorithm, where
$n$ is the number of vertices in $G$ and
$f$ is the feedback edge number of $G$.
Our algorithm thus runs in truly subquadratic time for digraphs
of $f=\bigO(n^{\frac{1}{3}-\epsilon})$ for any $\epsilon > 0$,
i.e., the number of edges is $n$ plus $\bigO(n^{\frac{1}{3}-\epsilon})$, and
is \emph{fully polynomial fixed parameter tractable},
the notion of which was first introduced by 
Fomin, Lokshtanov, Pilipczuk, Saurabh, and Wrochna
[ACM Trans.~Algorithms, 14(3):34:1--34:45, 2018].
We also show that the same result holds for vertex-weighted digraphs, where
the task is to compute the total weights of vertices reachable from each vertex.
\end{abstract}

\section{Introduction}
Consider the following problem concerning reachability on graphs.
Given a digraph $G$, count the number of vertices reachable from each vertex.
This problem is known by the name of \textsc{Descendant Counting}~\cite{cohen1997size}, and
it coincides with computing all the out-degrees of the transitive closure of $G$.
Computation of the size of the transitive closure has several applications, including
query optimization~\cite{lipton1995query,lipton1990practical},
sparse matrix multiplication~\cite{cohen1998structure}, and
social network analysis, wherein
it is used as a subroutine in
identifying the most influential set of individuals in a social network~\cite{chen2009efficient,kimura2010extracting,ohsaka2014fast}.

The current (theoretically) fastest algorithms for \textsc{Descendant Counting} have (at least) \emph{quadratic time} complexity:
they explicitly construct the transitive closure of $G$
in $\bigO(nm)$ time~\cite{purdom1970transitive,ebert81sensitive} by running a breadth-first search or
in $\bigO\left(nm \frac{\log n^2/m}{\log^2 n} + n^2\right)$ time~\cite{blelloch2008new} by using a sophisticated data structure for sparse graphs;
for dense graphs, this can be done in $ \tilde{\bigO}(n^\omega)$
time\footnote{$\tilde{\bigO}(g)$ denotes $\bigO(g \log^c g)$ for some positive integer $c$.}~\cite{adleman1978improved}
through fast matrix multiplication.
Here, $n$ is the number of vertices in $G$,
$m$ is the number of edges in the $G$, and
$\omega < 2.3728639 $~\cite{le2014powers} is the exponent of matrix multiplication.
Note that 
Cohen's celebrated approximation algorithm~\cite{cohen1997size}
estimates the number of reachable vertices
within a factor of $(1 \pm \epsilon)$ with high probability and
runs in $\bigO(\epsilon^{-2} n\log n)$ time, which is \emph{almost linear}.
Unfortunately, it has been proven by Borassi~\cite{borassi2016note} that
any exact algorithm that runs in \emph{truly subquadratic time},
i.e., in $\bigO(n^{2-\epsilon})$ time for any $\epsilon > 0$,
refutes the Strong Exponential Time Hypothesis (SETH)~\cite{impagliazzo2001complexity}.
The SETH states that for any $\epsilon > 0$
there exists some integer $k \geq 3$ such that \textsc{$k$-Satisfiability} on $n$ variables cannot be solved in $\bigO(2^{(1-\epsilon)n})$ time.
In particular, the same result is still true for sparse acyclic digraphs ($m=\bigO(n)$).

Nevertheless, in this study, we quest for truly subquadratic time \emph{as well as} exact algorithms for \textsc{Descendant Counting}.
To circumvent the quadratic time barrier,
we follow the framework of parameterized algorithms.
Given a parameter $k$ in addition to the input size,
a problem is referred to as \emph{fixed parameter tractable (FPT)}
if it is solvable in $g(k) \cdot |I|^{\bigO(1)}$ time,
where $g$ is some computable function depending only on parameter $k$ and
$|I|$ is the input size, e.g., $|I|=n+m$ in our case.
While FPT algorithms have been actively studied for \emph{NP-hard} problems 
(see, e.g., \cite{cygan2015parameterized}),
the concept of ``FPT inside P'' has opened up
a new exciting line of research~\cite{giannopoulou2017polynomial}.

Our contribution is that we present
an exact parameterized algorithm for
\textsc{Descendant Counting} that has running time $\bigO(f^3n)$,
where $n$ is the number of vertices in $G$ and
$f$ is the feedback edge number of $G$.
The \emph{feedback edge number} is the minimum number of edges,
the removal of which renders the underlying undirected graph acyclic;
this parameter has been used to develop parameterized algorithms for graph problems in P, e.g.,
\textsc{Maximum Matching} \cite{mertzios2020power},
\textsc{Betweenness Centrality} \cite{bentert2018adaptive},
\textsc{Hyperbolicity} \cite{fluschnik2019when},
\textsc{Triangle Listing} \cite{bentert2019parameterized}, and
\textsc{Diameter} \cite{bentert2019parameterizeda}.
Hence, for ``very tree-like'' digraphs having $f = \bigO(n^{\frac{1}{3}-\epsilon})$
for any $\epsilon > 0$,
i.e., the number of edges is bounded by $m = n + \bigO(n^{\frac{1}{3}-\epsilon})$,
our algorithm runs in
$\bigO(n^{2-3\epsilon})$---truly subquadratic---time
and thus outperforms the current fastest algorithms described above.\footnote{
Even when $G$ is a polytree, i.e., $m=n-1$,
the naive algorithms show quadratic time complexity,
because the transitive closure of $G$ can be of size $\bigO(n^2)$.
In contrast,
our algorithm no longer constructs the transitive closure.
}
On the other hand,
if it holds that $f = \Omega(n^{\frac{1}{3}+\epsilon})$,
which would be the case for real-world networks,
the proposed algorithm requires more than $\bigO(nm)$ time.
Furthermore, the dependence of the time complexity on parameter $f$ is \emph{polynomial}; such an algorithm,
introduced by Fomin, Lokshtanov, Pilipczuk, Saurabh, and Wrochna~\cite{fomin2018fully},
is called \emph{fully polynomial FPT}.
We also show that the same result holds for vertex-weighted digraphs, where the task is to compute 
the total \emph{weights} of vertices reachable from each vertex.

We here stress that
some graph parameters do not admit truly subquadratic time, fully polynomial FPT algorithms for \textsc{Descendant Counting}: 
Ogasawara~\cite{ogasawara2018fully} showed that
under the SETH,
a $k^{\bigO(1)} n^{2-\epsilon}$-time algorithm does not exist for any $\epsilon > 0$,
where $k$ denotes the \emph{treewidth} of $G$;
the same hardness applies to the case where $k$ is the \emph{feedback vertex number} of $G$,
which is the minimum number of vertices the removal of which renders the underlying undirected graph acyclic,
because Ogasawara used Borassi \cite{borassi2016note}'s reduction which constructs a digraph whose feedback vertex number is $\bigO(\log n)$.

\section{Preliminaries}
\paragraph{Notations and Definitions.}
For a digraph $G = (V,E)$,
let $V(G)$ and $E(G)$ denote the vertex set $V$ and the edge set $E$ of $G$, respectively.
Throughout this paper,
all the digraphs are simple; i.e., they have no self-loops and no multi-edges.
For a subset of vertices $S \subseteq V(G)$,
the subgraph induced by $S$ is denoted by $G[S]$.
A digraph is said to be \emph{acyclic}
if it contains no directed cycles,
to be \emph{weakly connected} if the underlying undirected graph is connected, and
to be \emph{strongly connected} if every vertex can reach every other vertex.
A \emph{polyforest} is a digraph, the underlying undirected graph of which is a forest.
A \emph{path} $P := (v_0, v_1, \ldots v_\ell)$ is a digraph with
vertex set $V(P) := \{ v_0, v_1, \ldots, v_\ell \}$ and
edge set $E(P) := \{ (v_0, v_1), (v_1, v_2), \ldots, (v_{\ell-1}, v_\ell) \}$,
where $v_0, v_1, \ldots v_\ell$ are distinct.
For a digraph $G$ and for an edge $(u,v)$ and a path $P$,
we denote
$G+(u,v) := (V(G), E(G) \cup \{(u,v)\})$ and $G+P := (V(G) \cup V(P), E(G) \cup E(P))$.
Given a vertex weighting $\mathbf{a}: V(G) \to \mathbb{R}$,
we denote by $a_v$ the weight for vertex $v$ and
abuse notation by writing
$ \mathbf{a}(S) = \sum_{v \in S} a_v $ for vertex set $S \subseteq V(G)$.

For a digraph $G$, the \emph{reachability set} of vertex $v$, denoted  $R_G(v)$,
is defined as the set of vertices reachable from $v$ on $G$ (including $v$ itself), and
the \emph{reachability number} of vertex $v$ is defined as $r_G(v) = |R_G(v)|$.
For a vertex weighting $\mathbf{a}$,
the \emph{weighted reachability number} of vertex $v$ is defined as
$r_{G,\mathbf{a}}(v) = \mathbf{a}(R_G(v))$. 
We formally define the descendant counting problem and its vertex-weighted version below.

\begin{problem}[\textsc{Descendant Counting}]
    Given a digraph $G$,
    the task is to compute
    the reachability number $r_G(v)$ for each vertex $v$ in $G$.
\end{problem}

\begin{problem}[\textsc{Weighted Descendant Counting}]
    Given a digraph $G$ and a vertex weighting $\mathbf{a}: V(G) \to \mathbb{R}$,
    the task is to compute
    the weighted reachability number $r_{G,\mathbf{a}}(v)$ for each vertex $v$ in $G$.
    In particular, the case where $a_v=1$ for all $v \in V(G)$ corresponds to \textsc{Descendant Counting}.
\end{problem}

A \emph{feedback edge set} in a digraph is a set of edges,
the deletion of which renders the digraph a polyforest.
The \emph{feedback edge number} is defined as
the minimum size of any feedback edge set.
That is, the feedback edge number of $G$
is equal to $ |E(G)|-|V(G)|+c $, where
$c$ is the number of weakly connected components in $G$,
which takes 1 if $G$ is entirely weakly connected.

\begin{remark}
    Our definition of a feedback edge set coincides with that for an undirected graph.
    For a digraph, the feedback arc set, the deletion of which renders the digraph acyclic, is usually adopted.
    We adopt the present definition, because
    solving \textsc{Descendant Counting} in truly subquadratic time for acyclic digraphs still falsifies the SETH.
\end{remark}

The \emph{condensation} $ G^\SCC $ of a digraph $G$
is defined as the digraph obtained from $G$ by
contracting each strongly connected component.
Formally, the vertices in $ G^\SCC $
are the strongly connected components in $G$, and
there exists an edge from a vertex $C$ to another vertex $C'$ in $G^\SCC$
if and only if
there exists an edge $(u,v) \in E(G)$
such that $u \in C, v \in C'$.
Note that the condensation is acyclic.
Let $\pi : V(G) \to V(G^\SCC)$ denote a mapping from vertices in $G$ to vertices in $G^\SCC$;
i.e., $\pi(v) = C $ whenever $v \in C \in V(G^\SCC)$.
We abuse notation by writing
$\pi((u,v)) = (\pi(u), \pi(v))$ for edge $(u,v)$ and
$\pi(S) = \{\pi(e) \mid e \in S\}$ for set $S$.

\paragraph{Warm-up: Linear-time Algorithm for a Polyforest.}
Let us first consider the case where the input graph is a polyforest $T$.
Noting that the weighted reachability number $r_{T,\mathbf{a}}(v)$ of vertex $v$ is
the sum of the weighted reachability numbers over its out-neighbors plus $a_v$,
i.e.,
\begin{align}
    r_{T,\mathbf{a}}(v) = a_v + \sum_{w : (v,w) \in E(T)} r_{T,\mathbf{a}}(w),
\end{align}
we are able to determine $r_{T,\mathbf{a}}(v)$ in a bottom-up fashion.
\cref{alg:polytree} shows the precise pseudocode.
Because the topological ordering can be found in linear time~\cite{tarjan1976edge},
\cref{alg:polytree} runs in $\bigO(n)$ time.
This linear-time algorithm is used as a subroutine,
as described in the following section.

\begin{algorithm}[tbp]
\caption{$\bigO(n)$-time algorithm for a polyforest.}
\label{alg:polytree}
\begin{algorithmic}[1]
\Require polyforest $T$ and vertex weighting $\mathbf{a}$.
\State find a topological ordering of $V(T)$.
\ForAll{$v \in V(T)$ in reverse topological order}
    \State $r(v) \leftarrow a_v + \sum_{w : (v,w) \in E(T)} r(w)$.
\EndFor
\State \Return $r$.
\end{algorithmic}
\end{algorithm}

\section{Fully Polynomial FPT Algorithm for Bounded Feedback Edge Number}
We now consider a general digraph of feedback edge number $f$.
Unlike in the case of a polyforest,
reachability numbers cannot be written as the sum of reachability numbers,
because out-neighbors' reachability sets can \emph{overlap} each other.
Our strategy is based on incremental update; i.e.,
(1) we first delete $f$ edges and
solve \textsc{(Weighted) Descendant Counting} on the resulting polyforest, and
(2) we then revert the $f$ edges serially and update the reachability number.

\subsection{Efficient Incremental Update on Acyclic Digraphs}
Let $G$ be an acyclic digraph of feedback edge number $f$ and $(s,t) \in V(G) \times V(G)$ be an edge not in $G$.
Assume that inserting $(s,t)$ does not render $G$ cyclic.
Given the reachability number $r_G$ for $G$,
we obtain the reachability number $r_{G+(s,t)}$ for $G+(s,t)$ as follows.
Let $\RU$ denote the set of vertices that can reach $s$ on $G$ and
$\RD$ denote the set of vertices reachable from $t$ on $G$.
Here, $\RU$ and $\RD$ are disjoint as $G + (s,t)$ is acyclic.
We also remark that
only vertices in $\RU$ can \emph{newly} reach some vertices in $\RD$.
Obviously, we can update reachability numbers for such vertices by
merely running a breadth-first search,
which, however, consumes quadratic time in the worst case.

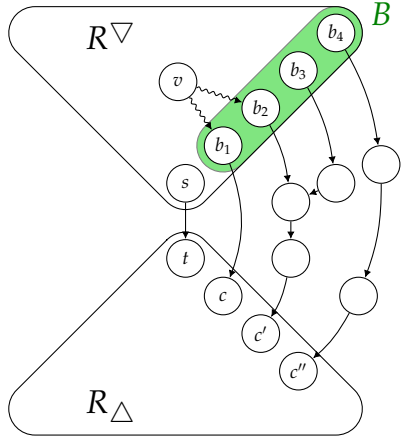
\begin{figure}[tbp]
\centering
\scalebox{0.5}{
\begin{tikzpicture}

\node[mynode](s)  at (0,1) {$s$};
\node[mynode](b1) at (1,2) {$b_1$};
\node[mynode](b2) at (2,3) {$b_2$};
\node[mynode](bi) at (3,4) {$\cdots$};
\node[mynode](bf) at (4,5) {$b_f$};
\node[mynode](v)  at (-0.2,3.7) {$v$};

\node[mynode](t)  at (0,-1) {$t$};
\node[mynode](c1) at (1,-2) {$c$};
\node[mynode](c2) at (2,-3) {$c'$};
\node[mynode](c3) at (3,-4) {$c''$};

\node[mynode] at (-4,5) (ul)  {};
\node[mynode] at (-4,-5) (ll) {};
\node[mynode] at (4,-5) (lr)  {};

\node[mynode](i2) at (2.8,0.5) {};
\node[mynode](i3) at (4,1)     {};
\node[mynode](i4) at (5.2,1.5) {};
\node[mynode](i5) at (2.8,-1)  {};
\node[mynode](i6) at (4.6,-2)  {};

\draw[thick,black,fill=white] \convexpath{s,ul,bf}{0.7cm};
\draw[thick,black,fill=white] \convexpath{t,lr,ll}{0.7cm};
\draw[thick,fill=green!80!black,opacity=0.5] \convexpath{b1,bf}{0.7cm};

\node[draw=none,fill=none,font=\Huge] at (-2,5) {$\RU$};
\node[draw=none,fill=none,font=\Huge] at (-2,-5) {$\RD$};
\node[draw=none,fill=none,font=\Huge,green!50!black] at (5.2,5.5) {$B$};

\node[mynode](s) at (0,1) {$s$};
\node[mynode](b1) at (1,2) {$b_1$};
\node[mynode](b2) at (2,3) {$b_2$};
\node[mynode](bi) at (3,4) {$b_3$};
\node[mynode](bf) at (4,5) {$b_4$};
\node[mynode](v) at (-0.2,3.7) {$v$};

\node[mynode](t) at (0,-1) {$t$};
\node[mynode](c1) at (1,-2) {$c$};
\node[mynode](c2) at (2,-3) {$c'$};
\node[mynode](c3) at (3,-4) {$c''$};

\draw[myedge] (s)--(t);
\draw[myedge] (b1) to [bend left=20] (c1);
\draw[myedge] (b2) to [bend left=10] (i2);
\draw[myedge] (bi) to [bend left=10] (i3);
\draw[myedge] (bf) to [bend left=10] (i4);
\draw[myedge] (i3) to                (i2);
\draw[myedge] (i2) to                (i5);
\draw[myedge] (i4) to [bend left=10] (i6);
\draw[myedge] (i5) to [bend left=10] (c2);
\draw[myedge] (i6) to [bend left=10] (c3);

\draw[myedge,mysnake] (v) to (b2);
\draw[myedge,mysnake] (v) to (b1);

\end{tikzpicture}
}

\caption{Illustration of $\RU$, $\RD$, and $B$.
Paths from $b_1, b_2, b_3, b_4$ to $s$ and 
paths from $t$ to $c,c',c''$ are omitted for simplicity.
Each vertex in $B$ can reach $\RD$ without passing through $(s,t)$.
Because vertex $v$ can reach $b_1$ and $b_2$ and their reachable sets,
it can reach vertices of $\RD \setminus (R_G(b_1) \cup R_G(b_2))$ for the first time on $G+(s,t)$.
}
\label{fig:boundary-set}
\end{figure}

To bypass this inefficiency, further definitions are required.
We say that a vertex $b \in \RU$ is a \emph{boundary}
if there exists a path from $b$ to some vertex of $\RD$
that does not pass through $(s,t)$ and
the internal vertices of which do not touch $\RU$.
We define the \emph{boundary set} as the set of all boundaries
and denote it by $B \subseteq \RU$.
See \cref{fig:boundary-set}.
For a vertex $v \in \RU$, we call $R_G(v) \cap B$ the \emph{restricted boundary set} for $v$.
Observe that the set difference of $R_{G+(s,t)}(v)$ and $R_{G}(v)$ for $v \in V(G)$
can be expressed by using $\RD$ and $R_G(b)$'s for $b \in B$:
\begin{align}
\label{eq:equiv}
R_{G+(s,t)}(v) \setminus R_G(v) =
\begin{cases}
\RD \setminus \bigcup_{b \in R_G(v) \cap B} R_G(b) & \textit{if } v \in \RU, \\
\emptyset & \textit{otherwise.}
\end{cases}
\end{align}
It is determined that
$ r_{G+(s,t)}(v) = r_G(v) + |\RD \setminus \bigcup_{b \in R_G(v) \cap B} R_G(b)| $ for vertex $v \in \RU$.
The boundary set has the following convenient properties
to compute \cref{eq:equiv} efficiently,
whose proofs are deferred to \cref{subsec:proofs}.
\begin{lemma}
\label{lem:B-size}
Let $G$ be an acyclic digraph of feedback edge number $f$ and $(s,t)$ be an edge not in $G$, and assume $G+(s,t)$ acyclic.
Then, the boundary set $B$ has at most $f$ vertices;
i.e., $|B| \leq f$.
\end{lemma}

\begin{lemma}
\label{lem:gain-size}
Let $G$ be an acyclic digraph of feedback edge number $f$ and $(s,t)$ be an edge not in $G$, and assume $G+(s,t)$ acyclic.
Then, the collection of the restricted boundary set for $v \in \RU$
has at most $2f$ distinct sets; i.e.,
\begin{align}
    |\{ R_G(v) \cap B \mid v \in \RU \}| \leq 2f.
\end{align}
\end{lemma}
\cref{lem:B-size} tells us that 
we can explicitly compute the reachability set for all boundaries in $\bigO(f(n+m))$ time.
By \cref{lem:gain-size},
we have that the space to store
$\bigcup_{b \in R_G(v) \cap B} R_G(b)$ in \cref{eq:equiv} for all $v \in \RU$ is bounded by $2f \cdot |R_G(\cdot)| =  \bigO(f n)$.
As a by-product, we can say that
the collection of the set differences between $R_{G+(s,t)}(v)$ and $R_G(v)$
for all $v \in \RU$ are of cardinality at most $2f$;
i.e., $ |\{ R_{G+(s,t)}(v) \setminus R_G(v) \mid v \in \RU \}| \leq 2f $.

\subsection{Algorithm Description for Acyclic Digraphs}
\cref{alg:fe} shows the precise pseudocode of our algorithm that,
given an acyclic digraph $G$ of feedback edge number $f$ and a vertex weighting $\mathbf{a} : V(G) \to \mathbb{R}$,
computes the reachability number $r_{G,\mathbf{a}}$ for all vertices in $G$.
We first construct a polyforest $T$ on $V(G)$ by deleting $f$ edges from $G$ and invoke \cref{alg:polytree} on $T$ with $\mathbf{a}$ to obtain its weighted reachability number $r_{T,\mathbf{a}}$.
Let $E(G) \setminus E(T) = \{(s_1,t_1), \ldots, (s_f,t_f)\}$ be the $f$ edges to be reverted to $T$ in an arbitrary order.
Let $G^{(0)} = T$ and define $G^{(i)} = G^{(i-1)} + (s_i,t_i) $ for each $i \in \{1, \ldots, f\}$.
For each $i \in \{0,1,\ldots, f\}$,
let $r^{(i)}$ be the weighted reachability number for $G^{(i)}$;
in particular, we have that $r^{(0)} = r_{T,\mathbf{a}}$.

The remaining part of the algorithm consists of $f$ rounds, which reverts each of the $f$ edges serially.
Given the current $r^{(i-1)}$ at the beginning of the $i$-th round with $i \in \{1,\ldots,f\}$,
we calculate $r^{(i)}$ as follows.
As in the previous section, let $\RU$ be the set of vertices that can reach $s_i$ on $G^{(i-1)}$ and
$\RD$ the set of vertices reachable from $t_i$ on $G^{(i-1)}$.
We compute the boundary set $B$ with regard to $G^{(i-1)}$ and $(s_i, t_i)$.
This computation can be done by
running a breadth-first search starting from $\RD$
that does not go forward whenever it touches $\RU$
on the transposed counterpart of $G^{(i-1)}$.
Then, for each boundary $b \in B$,
we mark vertices in $\RU$ that can reach $b$ and
compute the reachability set of $b$ in $G^{(i-1)}$, denoted by $R(b)$.
Let $\texttt{MARK}[v]$ be the set of boundaries marked for $v$; i.e., $ \texttt{MARK}[v] = R_{G^{(i-1)}}(v) \cap B $.
We know that for every $v \in \RU$,
$ R_{G^{(i)}}(v) \setminus R_{G^{(i-1)}}(v) $
is equal to
$\RD \setminus \cup_{b \in \texttt{MARK}[v]} R(b)$ because of \cref{eq:equiv}.
Hence, we declare an empty trie $\texttt{GAIN}$ for
storing $ \mathbf{a}(\RD \setminus \cup_{b \in \texttt{MARK}[v]} R(b)) $ with key $\texttt{MARK}[v]$ for every $v \in \RU$.
It should be noted that
we can omit the reachability set computation for some vertices, because
$ \texttt{MARK}[u] = \texttt{MARK}[v] $ may hold for $u \neq v$.
Finally, we compute $r^{(i)}(v)$ as
$ r^{(i-1)}(v) + \texttt{GAIN}[\texttt{MARK}[v]]$ for $ v \in \RU$ and
$ r^{(i-1)}(v) $ for $ v \in V(G) \setminus \RU$ and
insert $(s,t)$ into $G^{(i-1)}$ to obtain $G^{(i)}$.
Having completed the $f$ rounds, we return $r^{(f)}$.

\begin{algorithm}[tbp]
\caption{$\bigO(f^3 n + f^2 m)$-time algorithm for
an acyclic digraph of feedback edge number $f$.}
\label{alg:fe}
\begin{algorithmic}[1]
\Require acyclic digraph $G$ of feedback edge number $f$ and vertex weighting $\mathbf{a}$.
\State compute a polyforest $T$ on $V(G)$, the edge set of which is obtained by removing $f$ edges from $G$. \label{linum:fe:polytree}
\State invoke \cref{alg:polytree} on $T$ with $\mathbf{a}$ to obtain $r_{T,\mathbf{a}}$. \label{linum:fe:rT}
\State let $G^{(0)} \leftarrow T$ and $r^{(0)} \leftarrow r_{T,\mathbf{a}}$.
\State let $E(G) \setminus E(T) = \{ (s_1, t_1), \ldots, (s_f, t_f) \}$ \emph{in any order}.
\For{$i = 1 \To f$}
    \State compute set $\RU$ of vertices that can reach $s_i$ on $G^{(i-1)}$. \label{linum:fe:RU}
    \State compute set $\RD$ of vertices reachable from $t_i$ on $G^{(i-1)}$. \label{linum:fe:RD}
    \State compute boundary set $B \subseteq \RU$. \label{linum:fe:B}
    \State declare empty set $\texttt{MARK}[v]$ \textbf{for all} $v \in \RU$.
    \ForAll{$b \in B$} \label{linum:fe:MARK-begin}
        \ForAll{$v \in \RU$ that can reach $b$ on $G^{(i-1)}$}
            \State $ \texttt{MARK}[v] \leftarrow \texttt{MARK}[v] + b $.
        \EndFor
        \State $R(b) \leftarrow$ reachability set of $b$ w.r.t.~$G^{(i-1)}$.
    \EndFor \label{linum:fe:MARK-end}
    \State declare empty trie $\texttt{GAIN}$.
    \ForAll{$v \in \RU$} \label{linum:fe:GAIN-begin}
        \If{$\texttt{MARK}[v]$ is not found in $\texttt{GAIN}$}  \label{linum:fe:GAIN-search}
            \State $ \texttt{GAIN}[\texttt{MARK}[v]] \leftarrow \mathbf{a}(\RD \setminus \cup_{b \in \texttt{MARK}[v]} R(b)) $. \label{linum:fe:GAIN-if}
        \EndIf
    \EndFor
    \State $r^{(i)}(v) \leftarrow r^{(i-1)}(v) + \texttt{GAIN}[\texttt{MARK}[v]]$
        \textbf{for all} $ v \in \RU $. \label{linum:fe:r1}
    \State $r^{(i)}(v) \leftarrow r^{(i-1)}(v)$ \textbf{for all} $ v \in V(G) \setminus \RU $. \label{linum:fe:r2}
    \State $ G^{(i)} \leftarrow G^{(i-1)} + (s_i,t_i). $  \label{linum:fe:G}
\EndFor
\State \Return $r^{(f)}$.
\end{algorithmic}
\end{algorithm}

\subsection{Correctness and Time Complexity}
\label{subsec:proofs}
We now verify the correctness and the time complexity of \cref{alg:fe}.
To this end, we first validate \cref{eq:equiv}.

\begin{observation}
\label{obs:equiv}
Let $G$ be an acyclic digraph and $(s,t)$ be an edge not in $G$, and assume $G+(s,t)$ acyclic.
Then, for any vertex $v$, \cref{eq:equiv} is correct.
\end{observation}
\begin{proof}
The case where $v \not \in \RU$ is obvious;
we prove for $v \in \RU$.
We show that
a vertex $x \in V(G)$ is in the set on the left hand side in \cref{eq:equiv}
if and only if
$x$ is in the set on the right hand side in \cref{eq:equiv}.
First, assume that $x \in R_{G+(s,t)}(v) \setminus R_G(v)$.
We have that any path from $v$ to $x$ on $G+(s,t)$ must pass through $(s,t)$.
Obviously, $x \in \RD$.
We also have that $x \not \in R_G(b)$ for any $b \in R_G(v) \cap B$,
because otherwise $v$ can reach $x$ via such $b$ without passing through $(s,t)$,
which results in a contradiction.
Consequently, $x \in \RD \setminus \bigcup_{b \in R_G(v) \cap B} R_G(b)$.

Now, assume that $x \in \RD \setminus \bigcup_{b \in R_G(v) \cap B} R_G(b)$.
We have that $x$ is reachable from $t$ (on $G$).
Thus, $v$ can reach $x$ on $G+(s,t)$ by passing through $(s,t)$,
i.e., $x \in R_{G+(s,t)}(v)$.
On the other hand, $v$ cannot reach $x$ on $G$,
because to exit $\RU$ without passing through $(s,t)$,
a path starting from $v$ must touch some $b \in R_G(v) \cap B$;
i.e., $x \not \in R_G(v)$.
Consequently, $x \in R_{G+(s,t)}(v) \setminus R_G(v)$, which completes the proof.
\end{proof}

We now prove the two lemmas on the boundary set.

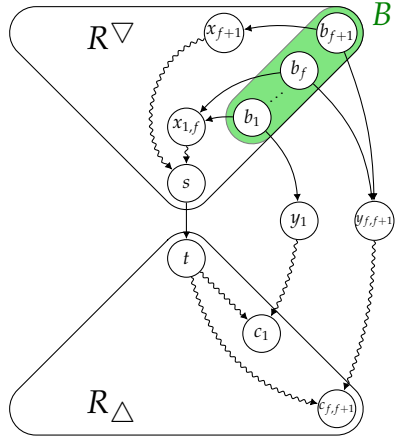
\begin{figure}[tbp]
\centering
\scalebox{0.5}{
\begin{tikzpicture}

\node[mynode](s) at (0,1) {$s$};
\node[mynode](b1) at (1.75,2.75) {$b_1$};
\node(bi) at (2.4,3.4) {\reflectbox{$\ddots$}};
\node[mynode](bf) at (3,4) {$b_f$};
\node[mynode](bf1) at (4,5) {$b_{f+1}$};
\node[mynode](x1) at (0,2.5) {$x_{1,f}$};
\node[mynode](x2) at (1,5) {$x_{f+1}$};

\node[mynode](t) at (0,-1) {$t$};
\node[mynode](c1) at (2,-3) {$c_1$};
\node[mynode,font=\normalsize](c2) at (4,-5) {$c_{f,f+1}$};

\node[mynode](y1) at (3,0) {$y_1$};
\node[mynode,font=\normalsize](y2) at (5,0) {$y_{f,f+1}$};

\node[mynode] at (-4,5) (ul) {};
\node[mynode] at (-4,-5) (ll) {};
\node[mynode] at (4,-5) (lr) {};

\draw[thick,black,fill=white] \convexpath{s,ul,bf1}{0.7cm};
\draw[thick,black,fill=white] \convexpath{t,lr,ll}{0.7cm};
\draw[thick,fill=green!80!black,opacity=0.5] \convexpath{b1,bf1}{0.7cm};

\node[draw=none,fill=none,font=\Huge] at (-2,5) {$\RU$};
\node[draw=none,fill=none,font=\Huge] at (-2,-5) {$\RD$};
\node[draw=none,fill=none,font=\Huge,green!50!black] at (5.2,5.5) {$B$};

\node[mynode](s) at (0,1) {$s$};
\node[mynode](b1) at (1.75,2.75) {$b_1$};
\node(bi) at (2.4,3.4) {\reflectbox{$\ddots$}};
\node[mynode](bf) at (3,4) {$b_f$};
\node[mynode](bf1) at (4,5) {$b_{f+1}$};
\node[mynode](x1) at (0,2.5) {$x_{1,f}$};
\node[mynode](x2) at (1,5) {$x_{f+1}$};

\node[mynode](t) at (0,-1) {$t$};
\node[mynode](c1) at (2,-3) {$c_1$};
\node[mynode,font=\normalsize](c2) at (4,-5) {$c_{f,f+1}$};

\draw[myedge] (s)--(t);
\draw[myedge] (b1) to [bend left=20] (y1);
\draw[myedge] (bf) to [bend left=20] (y2);
\draw[myedge] (bf1) to [bend left=10] (y2);
\draw[myedge] (b1) to [bend right=10] (x1);
\draw[myedge] (bf) to [bend right=20] (x1);
\draw[myedge] (bf1) to [bend right=10] (x2);

\draw[myedge,mysnake] (x1) to (s);
\draw[myedge,mysnake,bend right=60] (x2) to (s);
\draw[myedge,mysnake,bend left=10] (y1) to (c1);
\draw[myedge,mysnake,bend left=10] (y2) to (c2);
\draw[myedge,mysnake] (t) to (c1);
\draw[myedge,mysnake,bend right=30] (t) to (c2);

\end{tikzpicture}
}

\caption{Construction of
$G_0, G_1, \ldots, G_{f+1}$ in the proof of \cref{lem:B-size}.
The addition of three paths
$P_{b_i s}$, $P_{b_i c_i}$, and $P_{t c_i}$ 
increases the feedback edge number by at least one,
because $(b_i, x_i)$ and $(b_i, y_i)$ have thus far never been added.
(It could be the case that $x_i = b_j$ for some $j < i$.)
Consequently, $G_{f+1}$ has a feedback edge number of at least $f+1$, which is a contradiction.
}
\label{fig:constr}
\end{figure}

\begin{proof}[Proof of \cref{lem:B-size}]
We prove the statement by a contradiction.
Suppose that the boundary set $B$ contains at least $f+1$ vertices,
say, $b_1, b_2, \ldots, b_{f+1}$.
Without loss of generality, we can assume that
these vertices are sorted in reverse topological order,
so that $b_i$ cannot reach $b_j$ whenever $i < j$.

We construct a sequence of $f+2$ subgraphs of $G$,
denoted by $G_0, G_1, \ldots, G_{f+1}$, as follows.
The initial graph $G_0$ consists of a single edge $(s,t)$;
i.e., $G_0=(\{s,t\}, \{(s,t)\})$.
For each $i \in \{1,\ldots,f+1\}$,
$G_i$ is obtained from $G_{i-1}$ by adding the following three paths:
(1) a path $P_{b_i s}$ on $G[\RU]$ from $b_i$ to $s$,
(2) a path $P_{b_i c_i}$ from $b_i$ to some vertex $c_i \in \RD$
    the internal vertices of which belong to neither $\RU$ nor $\RD$, and
(3) a path $P_{t c_i}$ on $G[\RD]$ from $t$ to $c_i$.
See \cref{fig:constr}.
Note that the three paths are edge-disjoint, and
$f(G_i) = |E(G_i)|-|V(G_i)|+1$ for each $G_i$, where
$f(\cdot)$ denotes a feedback edge number, 
as it is weakly connected.

We now show that the feedback edge number increases by at least one from $G_{i-1}$ to $G_i$.
Let $x$ and $y$ be the vertex next to $b_i$
with regard to the two paths $P_{b_i s}$ and $P_{b_i c_i}$, respectively.
Then, we divide $P_{b_i s}$ into an edge $(b_i, x)$ and a subpath $P_{x s}$ from $x$ to $s$ and
divide $P_{b_i c_i}$ into an edge $(b_i, y)$ and a subpath $P_{y c_i}$ from $y$ to $c_i$.
See \cref{fig:constr}.
It is easy to see that
$G_{i-1}+P_{x s}+P_{t c_i}+P_{y c_i}$ is weakly connected.
Then, consider the addition of $(b_i, x)$ and $(b_i, y)$
to $G_{i-1}+P_{x s}+P_{t c_i}+P_{y c_i}$,
which yields $G_i$.
Observing that $x$ and $y$ have already been added and
$x \neq y$ (because $P_{b_i s}$ and $P_{b_i c_i}$ are edge-disjoint),
we can ensure that this addition preserves the weak connectivity and
increases the number of vertices by one for $b_i$ and the number of edges by two for $(b_i, x)$ and $(b_i, y)$;
i.e., $f(G_i) \geq f(G_{i-1}) + 1$.
Hence, $f(G_{f+1}) \geq f + 1 + f(G_0) = f+1$,
from which it follows that $f(G) \geq f(G_{f+1}) \geq f+1$, a contradiction.
\end{proof}

\begin{proof}[Proof of \cref{lem:gain-size}]
The proof is by contradiction.
Suppose that at least $2f+1$ vertices in $\RU$ have distinct restricted boundary sets;
i.e., $ |\{ R_G(v) \cap B \mid v \in \RU \}| \geq 2f+1 $.
We bound from below the number of edges in the induced subgraph $G[\RU]$.
It is easy to see that, if a vertex $v \in \RU \setminus B$ has an out-degree of one on $G[\RU]$, then both $v$ and $v$'s unique out-neighbor, say $w$, have 
exactly the same restricted boundary set;
i.e., $R_G(v) \cap B = R_G(w) \cap B$.\footnote{
Note that this observation does not hold if $v \in B$.
Suppose that $v \in B$ has an out-degree of one on $G[\RU]$, and
let $w$ be $v$'s unique out-neighbor.
We then have that
$R_G(v) \cap B \neq R_G(w) \cap B$ because
$v \in R_G(v) \cap B$ and $v \not \in R_G(w) \cap B$.
}
Hence,
there must be $2f+1-|B|$ vertices in $\RU \setminus B$ of an out-degree of at least two on $G[\RU]$
(because otherwise
$ |\{ R_G(v) \cap B \mid v \in \RU \}| \leq |\{ R_G(v) \cap B \mid v \in \RU \setminus B \}| + |B| \leq 2f $, which is a contradiction).
By \cref{lem:B-size}, we have that $2f+1-|B| \geq f+1$.
Observing that each vertex in $\RU-s$ has an out-degree of at least one on $G[\RU]$,
the number of edges in $G[\RU]$ is at least $(|\RU|-1) + (f+1) = |\RU|+f$.
Therefore,
the feedback edge number of $G[\RU]$ is $|E(G[\RU])|-|V(G[\RU])|+1 \geq |\RU|+f-|\RU|+1 = f+1$,
a contradiction.
\end{proof}

By \cref{lem:B-size,lem:gain-size},
we finally obtain the following:

\begin{theorem}
\label{thm:main}
Given an acyclic digraph $G$ and $\mathbf{a}: V(G) \to \mathbb{R}$,
\cref{alg:fe} correctly returns the weighted reachability number for all vertices in $G$ and runs in $\bigO(f^3n+f^2m)$ time and $\bigO(fn+m)$ space,
where $n=|V(G)|$, $m=|E(G)|$, and $f$ is the feedback edge number of $G$.
\end{theorem}
\begin{proof}
We first prove the correctness of \cref{alg:fe}.
To this end, we show that it holds that $r^{(i)} = r_{G^{(i)}, \mathbf{a}}$ for all $i \in \{0, 1, \ldots, f\}$ by induction on $i$.
The base case $i = 0$ is clear.
Now, assume that $r^{(i-1)} = r_{G^{(i-1)}, \mathbf{a}}$,
where $ i \in \{1, \ldots, f\} $.
In the $i$-th iteration,
$ \texttt{MARK}[v] $ is equal to $ R_{G^{(i-1)}}(v) \cap B $
for every $v \in \RU$.
Thus, by construction of $\texttt{GAIN}$, it holds that
$ r^{(i)}(v) = r^{(i-1)}(v) + \mathbf{a}( \RD \setminus \bigcup_{b \in R_{G^{(i)}}(v) \cap B} R_{G^{(i-1)}}(b) ) $
for every $v \in \RU$,
which is equal to $r_{G^{(i-1)},\mathbf{a}}(v) + \mathbf{a}(R_{G^{(i)}}(v) \setminus R_{G^{(i-1)}}(v)) = r_{G^{(i)},\mathbf{a}}(v)$ by the assumption and \cref{obs:equiv},
which completes the inductive step.

We now bound the running time of \cref{alg:fe}.
$T$ and $r_{T,\mathbf{a}}$ (steps~\ref{linum:fe:polytree}--\ref{linum:fe:rT}) can be constructed in linear time, because 
spanning tree computation and \cref{alg:polytree} are completed in linear time.
We now show that each of the $f$ rounds (steps~\ref{linum:fe:RU}--\ref{linum:fe:G}) consumes $\bigO(f^2n+fm)$ time.
First, we can compute $\RU, \RD$, and $B$ (steps~\ref{linum:fe:RU}--\ref{linum:fe:B}) in $\bigO(n+m)$ time
by running three breadth-first searches
starting from $s_i$, $t_i$, and $\RD$ on $G^{(i-1)}$, respectively.
By \cref{lem:B-size}, $B$ contains at most $f$ vertices.
Thus, $\texttt{MARK}$ can be constructed (steps~\ref{linum:fe:MARK-begin}--\ref{linum:fe:MARK-end}) in $\bigO(|B| \cdot (n+m)) = \bigO(f(n+m))$ time.
Note that each $\texttt{MARK}[v]$ contains at most $f$ vertices of $B$.
Then, observing that
there are at most $2f$ distinct sets in $\texttt{MARK}$ by \cref{lem:gain-size},
we can ensure that the algorithm reaches step~\ref{linum:fe:GAIN-if} at most $2f$ times and
compute the right hand side of step~\ref{linum:fe:GAIN-if} in $\bigO(fn)$ time by taking a union over (at most) $f$ sets.
Each search for $\texttt{GAIN}$ (step~\ref{linum:fe:GAIN-search}) and
each update of $\texttt{GAIN}$ (step~\ref{linum:fe:GAIN-if}) can be done in $\bigO(f)$ time
because the length of a binary representation of a key (i.e., any subset of $B$) is $\bigO(f)$.
It is thus determined that the construction of $\texttt{GAIN}$ (steps~\ref{linum:fe:GAIN-begin}--\ref{linum:fe:GAIN-if}) consumes $\bigO(f^2 n)$ time.
We can obviously update the reachability number and the digraph (steps~\ref{linum:fe:r1}--\ref{linum:fe:G}) in time $\bigO(fn)$.
Accordingly, the entire time complexity is bounded from above by $\bigO(f^3 n + f^2 m)$.
The space complexity is obvious.
\end{proof}

\subsection{Solution for General Digraphs}
We finally use \cref{alg:fe}
to solve \textsc{Weighted Descendant Counting} for arbitrary digraphs
of bounded feedback edge number.
For the sake of completeness, we prove the following observation.

\begin{observation}
\label{obs:scc}
If a digraph $G$ has a feedback edge number of at most $f$,
then so does the condensation $G^\SCC$.
\end{observation}
\begin{proof}
    Let $F$ be a feedback edge set of size (at most) $f$ of $G$.
    Then, the edge set $F' := \pi(F)$,
    which is of size at most $f$, is a feedback edge set of $G^\SCC$.
    This is because
    $G^\SCC - F'$ is equal to the graph $ (\pi(V(G)), \pi(E(G) - F)) $,
    which is a polyforest by definition of $F$.
\end{proof}

\begin{corollary}
For a digraph $G$ of feedback edge number $f$ and
a vertex weighting $\mathbf{a}: V(G) \to \mathbb{R}$,
\textsc{Weighted Descendant Counting} can be solved in $\bigO(f^3n)$ time and $\bigO(fn)$ space,
where $n = |V(G)|$.
\end{corollary}
\begin{proof}
Given $G$ and $\mathbf{a}$,
we first construct its condensation $G^\SCC$ in linear time~\cite{tarjan1972depth} and
a vertex weighting $\mathbf{b}: V(G^\SCC) \to \mathbb{R}$
such that $b_C = \mathbf{a}(C)$ for each strongly connected component $C$ of $G$.
We then invoke \cref{alg:fe} on $G^\SCC$ with $\mathbf{b}$,
which consumes at most $\bigO(f^3|V(G^\SCC)| + f^2|E(G^\SCC)|)$ time
because of \cref{thm:main,obs:scc}.
We finally compute the weighted reachability number for $G$ by
$r_{G, \mathbf{a}}(v) = r_{G^\SCC, \mathbf{b}}(\pi(v))$
for each
$v \in V(G)$,
where $\pi : V(G) \to V(G^\SCC)$ is a mapping from vertices in $G$ to vertices in $G^\SCC$.
The whole time complexity is thus bounded by $\bigO(f^3|V(G)|)$ since $|E(G)| \leq |V(G)| + f$.
The space complexity is obvious.
\end{proof}

\section*{Acknowledgments}
The author would like to thank Tomoaki Ogasawara for providing their master's thesis \cite{ogasawara2018fully} and the anonymous reviewers for their valuable comments and suggestions.

\bibliographystyle{alpha}
\bibliography{main}

\end{document}